\documentclass[default,iicol]{sn-jnl}% Default with double column layout

\usepackage{graphicx}%
\usepackage{multirow}%
\usepackage{amsmath,amssymb,amsfonts}%
\usepackage{amsthm}%
\usepackage{mathrsfs}%
\usepackage[title]{appendix}%
\usepackage{xcolor}%
\usepackage{textcomp}%
\usepackage{manyfoot}%
\usepackage{booktabs}%
\usepackage{algorithm}%
\usepackage{algorithmicx}%
\usepackage{algpseudocode}%
\usepackage{listings}%
\usepackage[numbers,sort&compress]{natbib}
\begin{document}

\title[Article Title]{Improved constraints for axion-like particles from 3-photon events at $e^+e^-$ colliders}


\author{\fnm{Aleksandr} \sur{Pustyntsev}}\email{apustynt@uni-mainz.de}

\author{\fnm{Marc} \sur{Vanderhaeghen}}\email{vandma00@uni-mainz.de}

\affil{Institut f\"ur Kernphysik and $\text{PRISMA}^+$ Cluster of Excellence, Johannes Gutenberg Universit\"at, D-55099 Mainz, Germany}


\abstract{Axions and axion-like particles (ALPs) are one of the most widely discussed extensions of the Standard Model when it comes to the strong CP problem and dark matter candidates. Current experiments are focused on the indirect searches of invisible pseudoscalars in a wide parameter range. In this paper we investigate limits on ALP mass, and its couplings to photons and leptons from 3-photon annihilation at $e^+e^-$ colliders. We provide detailed calculations and apply them to the particular kinematics of the Belle II experiment, covering the ALP mass range from few hundred MeV to around 10 GeV. Our results, which improve upon previous analyses by also including the ALP coupling to electrons, show that such future analyses will allow to significantly extend the ALP search range and impose much more stringent restrictions on their couplings.}

\maketitle

\section{Introduction}\label{sec1}

Initially proposed in 1977, the Peccei-Quinn theory so far is considered to be the most compelling strong CP problem resolution~\cite{Peccei:1977hh, Peccei:1977ur}. In this model a CP-violating phase is dynamically driven to zero, giving rise to a new pseudoscalar particle called axion~\cite{Weinberg:1977ma, Wilczek:1977pj}. A key property of the QCD axion is the linear proportionality between its couplings to the Standard Model particles and the axion mass.

During the last four decades numerous attempts have been made to find a signal of this particle, including both lab searches and astronomical observations~\cite{Graham:2015ouw, Irastorza:2018dyq}. Current constraints show that the QCD axion (in case it exists) must be very weakly interacting and thus is called "invisible", which forces one to concentrate on the possible indirect detection of this particle~\cite{Sikivie:2020zpn, Darme:2020sjf}.

In addition to the QCD axion mechanism, various Standard Model extensions with axion-like particles were proposed~\cite{Baker:2013zta,Bagger:1994hh,Arvanitaki:2009fg}. The main difference to the original model is that ALPs are not restricted to a linear mass-coupling relation. Furthermore, ALPs are considered to be promising dark matter candidates as being both very long-lived and weakly-interacting with the mass unconstrained by their interactions with other particles~\cite{Arias:2012az, Giannotti:2015kwo}. During the past few years there is a noticeable increase of interest in the MeV to GeV range~\cite{Dolan:2017osp,Millea:2015qra,Jaeckel:2015jla,Beacham:2019nyx,Gavela:2019cmq}. Although significant progress has been made recently~\cite{Agrawal:2021dbo,Antel:2023hkf}, there still remain untested regions in the parameter space, especially when compared to the low-mass region where astrophysical constraints can be applied.

The ALP-photon coupling has been the subject of numerous studies, approached from both theoretical~\cite{Dolan:2017osp,Mimasu:2014nea} and experimental perspectives~\cite{Belle-II:2020jti,BESIII:2022rzz}. However, these investigations often overlook the ALP-lepton coupling, despite the fact that it is much less constrained experimentally and thus warrants attention (see, for example, the recent work~\cite{Liu:2023bby}, where the significance of the ALP-photon and ALP-lepton couplings interplay is also emphasized).

At first glance, one might assume that additional interactions would amplify the cross section, thereby imposing even stricter bounds. However, this reasoning falls short when an extra decay channel significantly reduces the ALPs lifetime. Our goal is to refine current analyses by incorporating the ALP-electron interaction and assessing its potential impact on the constraints derived from electron-positron colliders.

In this work we investigate the mass and coupling constraints of ALPs in the MeV to GeV range from 3-photon events in $e^+e^-$ annihilation. We focus on the kinematical setting of the Belle II experiment. Section \ref{sec:2} provides a general overview of the given formalism with the discussion of couplings, matrix elements and cross sections. Section \ref{sec3} illustrates the main results and provides predictions for Belle II kinematics and constraints which follow from the calculated processes. Section \ref{sec4} summarizes our work.

\section{ALP formalism}\label{sec:2}
In this work we assume that ALPs in the MeV to GeV mass range couple predominantly to electroweak gauge bosons and electrons, i.e. decay only to visible states and are decoupled from the QCD sector. The last statement is justified by the fact that such coupling is strongly constrained by the searches of flavour-changing processes~\cite{Dolan:2014ska}.

The following section provides a short review of the relevant ALP interactions. The parameter space includes three variables - the ALP mass $m_a$ and its couplings to photons and electrons, which are denoted by $g_{a \gamma \gamma}$ and $g_{aee}$, respectively. 

In this paper, we do not focus on any specific UV-complete ALP model, and the exact origin of these couplings is unspecified. At the same time, our primary objective is to impose constraints on ALPs using Belle II data and it’s important to note that experimental setups can only detect effective couplings rather than tree-level ones~\cite{Liu:2023bby}. Therefore, the couplings $g_{a \gamma \gamma}$ and $g_{aee}$ must be considered as effective, encompassing both their tree-level values and potential loop-induced effects. 

In this framework we detail the calculations of ALP contributions to 2- and 3-photon annihilation of $e^+e^-$ pairs and the interplay of two couplings.

\subsection{Electroweak gauge invariance}\label{subsec:21}

The generic gauge-invariant Lagrangian of ALPs interaction with electroweak vector bosons has the form~\cite{Dolan:2017osp}

\begin{equation}\label{eq:ew}
    \mathcal{L}_{a-EW} = -\frac{g_{aBB}}{4}aB^{\mu \nu}\tilde{B}_{\mu \nu} -\frac{g_{aWW}}{4}aW_i^{\mu \nu}\tilde{W}_{i, \mu \nu},
\end{equation}
$a$ stands for the pseudoscalar ALP field, $B^{\mu \nu}$ and $W^{\mu \nu}_i$ refer to the $U\left(1\right)$ and $SU\left(2\right)$ field tensors, respectively. The corresponding dual pseudotensors are defined in the standard way, $\tilde{B}_{\mu \nu} = \frac{1}{2}\varepsilon_{\mu \nu \lambda \sigma} {B}^{ \lambda \sigma}$ and similarly for $W^{\mu \nu}_i$, while $g_{a BB}$ and $g_{a WW}$ are the coupling constants with dimension $\mbox{GeV}^{-1}$.

After symmetry breaking the effective interaction of ALPs and photons is given by

\begin{equation}
\mathcal{L}_{a \gamma \gamma} = -\frac{g_{a \gamma \gamma}}{4}aF^{\mu \nu}\tilde{F}_{\mu \nu},
\end{equation}
where the new coupling constant was defined

\begin{equation}
g_{a\gamma\gamma} = g_{aBB} \cos^2{\theta_W}+g_{aWW} \sin^2{\theta_W},
\end{equation}
with $\theta_W$ being the Weinberg angle.

Of course, the Lagrangian \eqref{eq:ew} also leads to a non-zero $aZ Z$,  $aWW$ and also  $a\gamma Z$ interactions. However, in the domain of interest corresponding to electron-positron collider with center-of-momentum energy below 10 GeV, these channels are irrelevant, since the production of $Z$ and $W$ bosons is prohibited by energy conservation and can safely be neglected. 

\begin{figure}[t]
\centering
\includegraphics{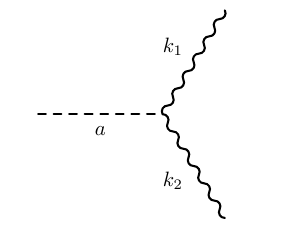} 
\caption{ALP decay to two photons in the lowest order.}
\label{fig:d1}
\end{figure}

The matrix element for the $a \to 2 \gamma$ decay shown in Fig. \ref{fig:d1} is given by

\begin{equation}\label{eq:ph}
\begin{split}
M_{a \to \gamma\gamma} \left(k_1, k_2\right) & = -ig_{a \gamma \gamma} k_{1,\kappa}k_{2,\beta} \varepsilon^{\kappa \beta \mu  \nu}\\
& \times \epsilon^*_{\mu}\left(k_1, \lambda_1\right)\epsilon^*_{\nu}\left(k_2, \lambda_2\right),
\end{split}
\end{equation}
where $\epsilon_{\mu}\left(k_1, \lambda_1\right)$ and $\epsilon_{\mu}\left(k_2, \lambda_2\right)$ are the polarization vectors of the photons with 4-momenta  $k_1$, $k_2$ and helicities $\lambda_1$, $\lambda_2$, respectively. Summing over the final helicities, we obtain

\begin{equation}
\sum_f \left| M_{a \to \gamma\gamma}\left(k_1, k_2\right) \right|^2 = 2 g^2_{a \gamma \gamma} \left(k_1k_2\right)^2.
\end{equation}

The corresponding decay width is then obtained as

\begin{equation}\label{eq:cs1}
\Gamma_{a \gamma \gamma} = \frac{g^2_{a \gamma \gamma}m_a^3}{64\pi}.
\end{equation}

\subsection{Interaction with leptons}\label{subsec:22}

We next discuss the ALP-fermion coupling and the corresponding decay rate. The generic interaction of ALPs with fermions is of the form

\begin{equation}
    \mathcal{L}_{aff} = -\frac{g_{aff}}{2m_f} \partial_{\mu} a  \Bar{f} \gamma^5 \gamma^{\mu} f,
\end{equation}
where $f$ stands for the fermion field, $m_f$ denotes its mass and $g_{aff}$ is the dimensionless coupling constant. 

From $\mathcal{L}_{aff}$ it is clear that lepton universality requires the large enhancement of ALP coupling to muon, namely

\begin{equation}
    g_{a \mu \mu} \approx \frac{m_{\mu}}{m_e}  g_{a e e}.
\end{equation}

In this paper we follow Alves and Wiener work~\cite{Alves:2017avw} and consider ALPs coupled only to electrons in order to avoid effects induced by this enhanced coupling, such as $\left(g-2\right)_{\mu}$ corrections on the muon anomalous magnetic moment.

At tree level $\mathcal{L}_{aff}$ can be equivalently reduced to a pseudoscalar coupling

\begin{equation}
\mathcal{L}_{aff}= -ig_{aff} a \Bar{f} \gamma^5 f.
\end{equation}

\begin{figure}[t]
\centering
\includegraphics{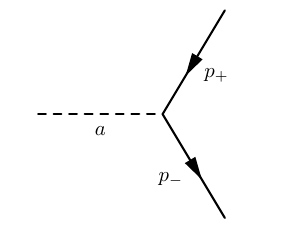} 
\caption{ALP decay to the lepton-antilepton pair in the lowest order.}
\label{fig:d2}
\end{figure}

We are interested in the $a \to e^+ e^-$ decay shown on Fig. \ref{fig:d2}, which has the amplitude

\begin{equation}
M_{a \to e^+ e^-} = g_{aee} \Bar{u} \left(p_-, s_-\right) \gamma^5 v\left(p_+, s_+\right), \\
\end{equation}
where $u\left(p_-, s_-\right)$ and $v\left(p_+, s_+\right)$ are the bispinors describing electron and positron with momenta $p_{\pm}$ and helicities $s_{\pm}$, respectively. 

In the domain of interest we can assume $m_e \ll m_a$. After summing over the final helicities, the squared amplitude for this process is given by

\begin{equation}
\sum_f \left|M_{a \to e^+ e^-} \right|^2 =2g_{aee}^2m_a^2.
\end{equation}

The corresponding decay width then has the form

\begin{equation}\label{eq:cs2}
\Gamma_{aee} =  \frac{g_{aee}^2m_a}{8\pi}.
\end{equation}

In the absence of interaction with other fields, the total ALP decay width is assumed to consist of two contributions

\begin{equation}
\Gamma_a = \Gamma_{aee} + \Gamma_{a \gamma \gamma}.
\end{equation}

\subsection{ALP production at $e^+e^-$ colliders}\label{subsec:23}

\begin{figure}[h]
\centering
\includegraphics{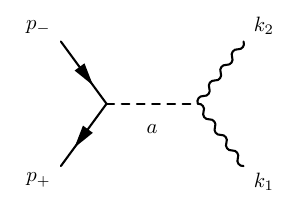} 
\caption{$e^+e^-$ annihilation into two photons through an intermediate ALP.}
\label{fig:d3}
\end{figure}

An ALP contributes to the 2-photon annihilation of $e^+e^-$ through the diagram shown on Fig. \ref{fig:d3}. The matrix element is

\begin{equation}
\begin{split}
M_{e^+e^- \to \gamma \gamma} & = ig_{aee}\frac{\Bar{v} \left(p_+, s_+\right)\gamma^5 u\left(p_-, s_-\right)}{s-m_a^2+im_a \Gamma_a} \\
& \times M_{a \to \gamma\gamma}\left(k_1,k_2\right),
\end{split}
\end{equation}
with $M_{a \to \gamma\gamma}\left(k_1,k_2\right)$ given in Eq. \eqref{eq:ph}.

As a function of $m_a$, this cross section is significantly different from zero only in a small region around $m_a^2=s=4E^2$, where $E$ denotes the initial electron (positron) energy in the center-of-momentum frame. Thus, for a fixed collider energy, the $e^+e^- \to \gamma \gamma$ process is not providing constraints on ALP parameters in a broad $m_a$, $g_{a \gamma \gamma}$, $g_{aee}$ parameter space in $e^+e^-$ annihilation. Therefore, in the following we investigate 3-photon final states.

\begin{figure}[h]
\centering
\includegraphics{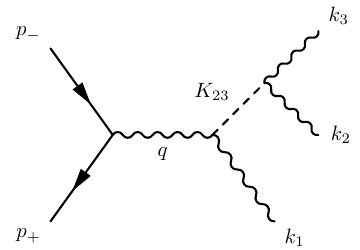} 
\caption{$e^+e^-$ annihilation into three photons involving the $g_{a\gamma\gamma}$ coupling. Graphs obtained from these by crossing are not shown, but are evaluated too.}
\label{fig:d4}
\end{figure}

Fig. \ref{fig:d4} shows the contribution of ALP-photon coupling resulting in 3-photon events. The corresponding amplitudes are given by

\begin{equation}
\begin{split}
& M_{e^+e^- \to \gamma \gamma \gamma \, \left(ALP_1\right)} = i\frac{H_{e^+e^- \to \gamma^* \to a\gamma} \left(k_1\right)}{K_{23}^2 - m_a^2 + i m_a \Gamma_a}\\
&  \times  M_{a \to \gamma\gamma}\left(k_2,k_3\right) + \mbox{crossed terms},
\end{split}
\end{equation}
where $H_{e^+e^-\to \gamma^* \to a\gamma}\left(k_i\right)$ stands for $e^+ e^- \to a \gamma_i$ amplitude

\begin{equation}
\begin{split}
H_{e^+e^- \to \gamma^* \to a\gamma} \left(k_i\right) & = -ieg_{a \gamma \gamma} \, \varepsilon_{\alpha \beta \mu \gamma } q^{\alpha} k_{i}^\beta\epsilon^{\gamma}\left(k_i, \lambda_i\right) \\
& \times \frac{ \Bar{v} \left(p_+, s_+\right)\gamma^{\mu} u\left(p_-, s_-\right)}{s} ,
\end{split}
\end{equation}
with $e$ representing the positron charge and $q=p_++p_-$ standing for the internal photon 4-momentum. We denote the ALP 4-momenta as $K_{23} =  k_2+k_3$.

It is generally assumed that ALPs are long-lived particles, i.e. their decay width $\Gamma_a$ is a small quantity, typically much smaller than the experimental resolution of the invariant mass of the two-photon system in which the ALP decays. Thus the integration over the phase space gives the main contribution only in a very small range of variables where the invariant mass of the photon pair produced by the ALP is close to $m_a^2$. 

In such kinematics the interference terms become unobservable and can be omitted. After the integration over the phase space the total cross section can be represented as a cross section obtained from only Feynman diagrams shown in Fig. \ref{fig:d4} multiplied by a factor of three to account for the 3 channels. Thus we obtain

\begin{equation}
\begin{split}
& \bar{\sum_i} \sum_f \left|H_{e^+e^- \to \gamma^*\to a\gamma} \left(k_1\right)\right|^2  \\
&= \frac{2e^2g^2_{a \gamma \gamma}}{s^2}  \left[\left(k_1p_+\right)^2+\left(k_1p_-\right)^2 \right] \left(p_-p_+\right),
\end{split}
\end{equation}
where $\bar{\sum_i} \sum_f$ denotes the average over initial helicities states and the sum over final. 

As we conclude this part, it’s important to note that the diagram in Fig. \ref{fig:d4} incorporates the ALP-photon coupling twice, but in distinct ways. Specifically, one vertex depicts a Primakoff-like process (with one photon being virtual), while the other corresponds with the ALP decay. Given that these couplings must be viewed as effective, they are differentiated by loop correction and, fundamentally, they are not identical, as emphasized in~\cite{Ferreira:2022xlw}. However, when the limit $m_e/m_a \ll 1$ is fulfilled, the difference becomes negligible, making the two effective couplings in good approximation equivalent once again. Further details can be found in Appendix \ref{appendix:a}.

\begin{figure*}
\centering
\includegraphics{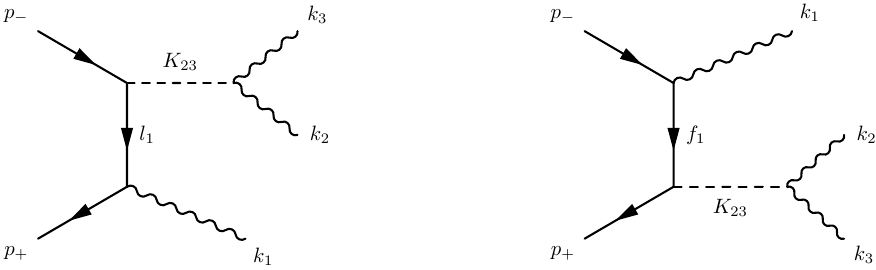}
\caption{$e^-e^+$ annihilation into three photons involving the $g_{aee}$ coupling. Graphs which are obtained by crossing are not shown, but are evaluated too.}
\label{fig:d5}
\end{figure*}

We next discuss the 3-photon production in $e^+ e^-$ annihilation which results from the ALP-electron coupling contribution, shown in Fig. \ref{fig:d5}. The corresponding amplitudes can be expressed as

\begin{equation}
\begin{split}
& M_{e^+e^- \to \gamma\gamma\gamma \, \left(ALP_2\right)} \\
& = i\frac{H_{e^+e^- \to a\gamma,1}\left(k_1\right) + H_{e^+e^- \to a\gamma,2}\left(k_1\right) }{K_{23}^2 - m_a^2 + i m_a \Gamma_a}\\
& \times  M_{a \to \gamma\gamma}\left(k_2,k_3\right) + \mbox{crossed terms},
\end{split}
\end{equation}
where $H_{e^+e^- \to a\gamma,j}\left(k_i\right)$ denote amplitudes for the corresponding $2 \to 2$ process

\begin{gather}
\begin{split} 
H_{e^+e^- \to a\gamma,1} \left(k_i\right) & = eg_{aee} \,\epsilon_{\eta}^*\left(k_i, \lambda_i\right) \\
& \times \Bar{v}\left(p_+, s_+\right) \gamma^{\eta} \frac{\hat{l}_i}{l_i^2} \gamma^{5} u \left(p_-, s_-\right)  ,
\end{split}\\
\begin{split}
H_{e^+e^- \to a\gamma,2}\left(k_i\right) & = eg_{aee} \, \epsilon_{\lambda}^*\left(k_i, \lambda_i\right) \\
& \times \Bar{v}\left(p_+, s_+\right) \gamma^5 \frac{\hat{f}_i}{f_i^2} \gamma^{\lambda} u \left(p_-, s_-\right)  , 
\end{split}
\end{gather}
with the internal electron momenta

\begin{equation}
l_i = k_i-p_+, \quad
f_i = p_- - k_i.
\end{equation}

It is worth noticing that there is no interference between this set of diagrams and the diagrams shown in Fig. \ref{fig:d4}. Using the same arguments as before, we conclude that for the cross section calculation, we only need to evaluate the two topologies shown in Fig. \ref{fig:d5}, as

\begin{figure*}[t]
\centering
\includegraphics[width=0.45\textwidth]{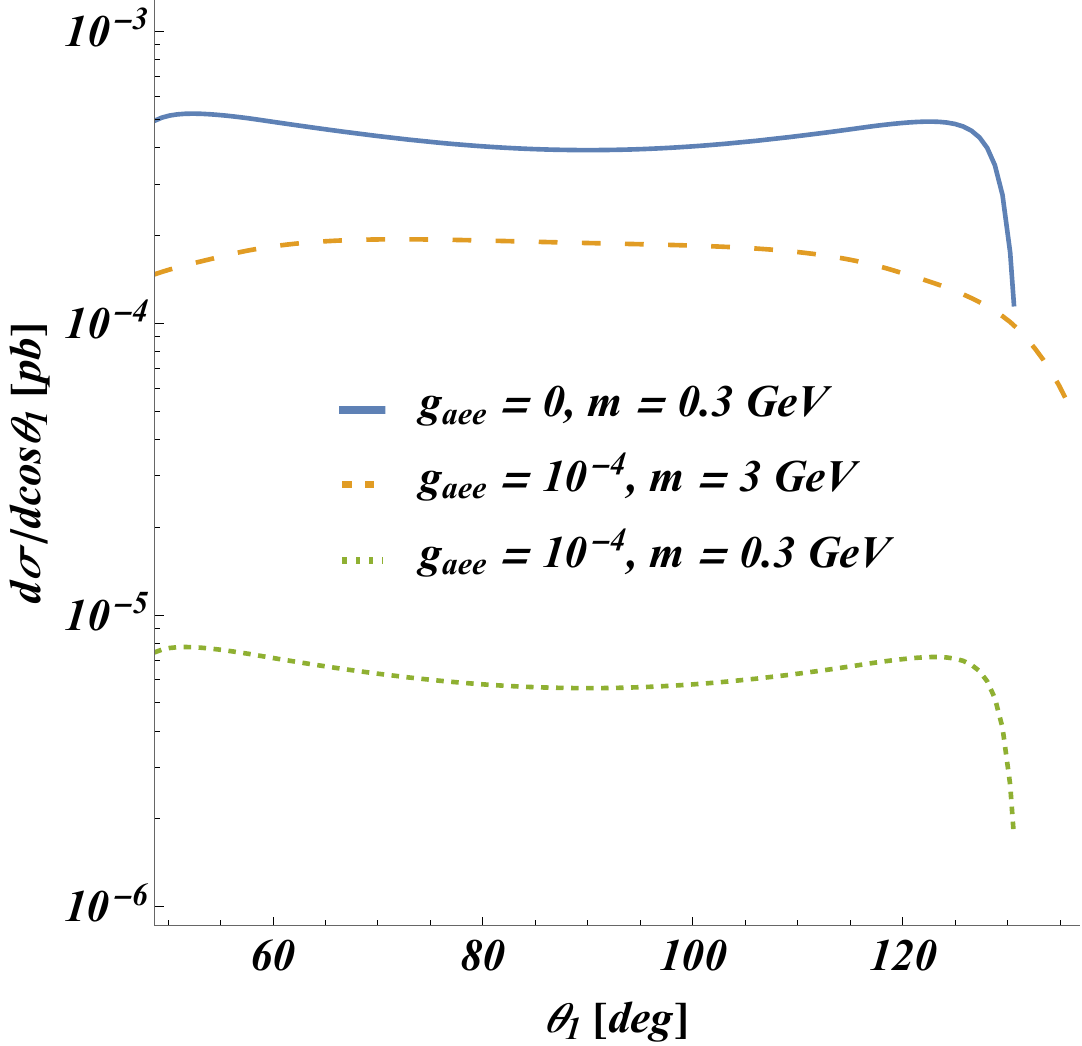} \quad \quad \includegraphics[width=0.45\textwidth]{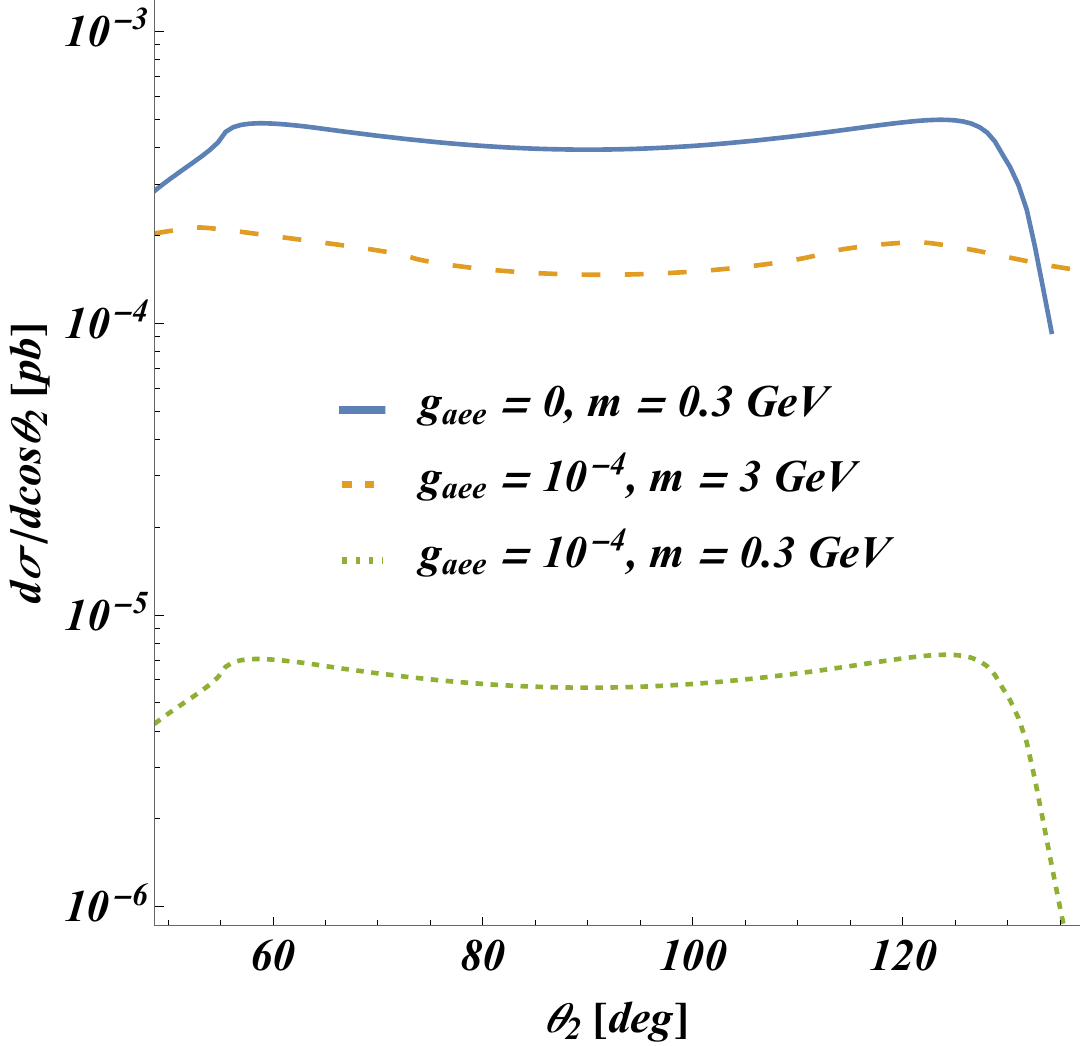} 
\caption{Angular distributions of $e^+e^- \to \gamma \gamma \gamma$ annihilation for a process with an intermediate ALP in Belle II kinematics for two values of $m_a$ and two values of $g_{aee}$. The value $g_{a \gamma \gamma} = 10^{-4} \, \mbox{GeV}^{-1}$ is used for all curves.}
\label{fig:d6}
\end{figure*}

\begin{equation}
\begin{split}
& \bar{\sum_i} \sum_f \left|H_{e^+e^- \to a\gamma,1}\left(k_1\right) + H_{e^+e^- \to a\gamma,2}\left(k_1\right)\right|^2  \\
& = e^2g_{aee}^2 \left(\frac{ p_-k_1}{p_+k_1} +\frac{ p_+k_1}{p_-k_1} + 2 \frac{\left( p_+ K_{23}\right)\left(p_-K_{23}\right)}{\left(p_-k_1\right)\left(p_+k_1\right)} \right).
\end{split}
\end{equation}

\subsection{Cross section and observables}\label{subsec:24}

The cross section of $e^+e^- \to \gamma\gamma\gamma$ process is given by the expression

\begin{eqnarray}
 \sigma &=& \frac{1}{3!} \int d_{LIPS} \left(2\pi\right)^4 \delta^{\left[4\right]}\left(p_- + p_+ - k_1 - k_2 - k_3\right) \nonumber \\
&\times&  \frac{1}{2s} \bar{\sum_i} \sum_f \left|M_{e^+e^- \to \gamma\gamma\gamma}\right|^2, 
\end{eqnarray}
where $d_{LIPS}$ stands for the Lorentz-invariant phase space of the three final photons

\begin{equation}
d_{LIPS} = \frac{d^3k_1}{2\omega_1 \left(2\pi\right)^3}\frac{d^3k_2}{2\omega_2 \left(2\pi\right)^3}\frac{d^3k_3}{2\omega_3 \left(2\pi\right)^3}.
\end{equation}

After the integration with the delta function, the phase space can be expressed as

\begin{equation}
\begin{split}
& \, \, d_{LIPS} \left(2\pi\right)^4 \delta^{\left[4\right]}\left(p_- + p_+ - k_1 - k_2 - k_3\right)  \\
& = \frac{1}{2^8 \pi^5} \frac{\omega_1 \omega_2}{2E + \omega_1 \left(\cos{\theta_{12}} - 1 \right)} d\omega_1  d\Omega_1 d\Omega_2,
\end{split}
\end{equation}
with $\theta_{12}$ denoting the angle between $\textbf{k}_1$ and $\textbf{k}_2$ momenta. The remaining phase space is parameterized as

\begin{equation}
d\Omega_1 d\Omega_2 = 2 \pi d\phi \, d\cos{\theta_{1-}}d\cos{\theta_{2-}},
\end{equation}
where $\theta_{i-}$ is the angle between $\textbf{p}_-$ and $\textbf{k}_i$, which leads to

\begin{equation}
\cos{\theta_{12}} = \sin{\theta_{1-}}\sin{\theta_{2-}} \cos{\phi} + \cos{\theta_{1-}}\cos{\theta_{2-}}.
\end{equation}

Furthermore, in the center-of-momentum frame it holds 

\begin{equation}
\begin{cases}
\omega_1 + \omega_2+\omega_3 = 2E,\\
\textbf{k}_1+\textbf{k}_2+\textbf{k}_3=0,
\end{cases}
\end{equation}
allowing to express $\omega_2$ as

\begin{equation}
\omega_2 = \frac{2E\left(E-\omega_1\right)}{2E + \omega_1\left(\cos{\theta_{12}}-1\right)}.
\end{equation}

For the ALP-associated process, the photon which is opposite to the ALP in center-of-momentum frame is denoted by $k_1$. In this case, we can remove the integration over $d\omega_1$ using the definition of the delta function

\begin{equation}
\begin{split}
& \frac{1}{\left(K_{23}^2-m_a^2\right)^2+\left(m_a \Gamma_a\right)^2} \to \frac{\pi}{m_a \Gamma_a} \delta\left(K_{23}^2-m_a^2\right) \\
& = \frac{\pi}{m_a \Gamma_a} \frac{1}{4E} \, \delta\left(\omega_1 - \frac{4E^2-m^2_a}{4E}\right).
\end{split}
\end{equation}

\begin{figure*}[t]
\centering
\includegraphics[width=0.45\textwidth]{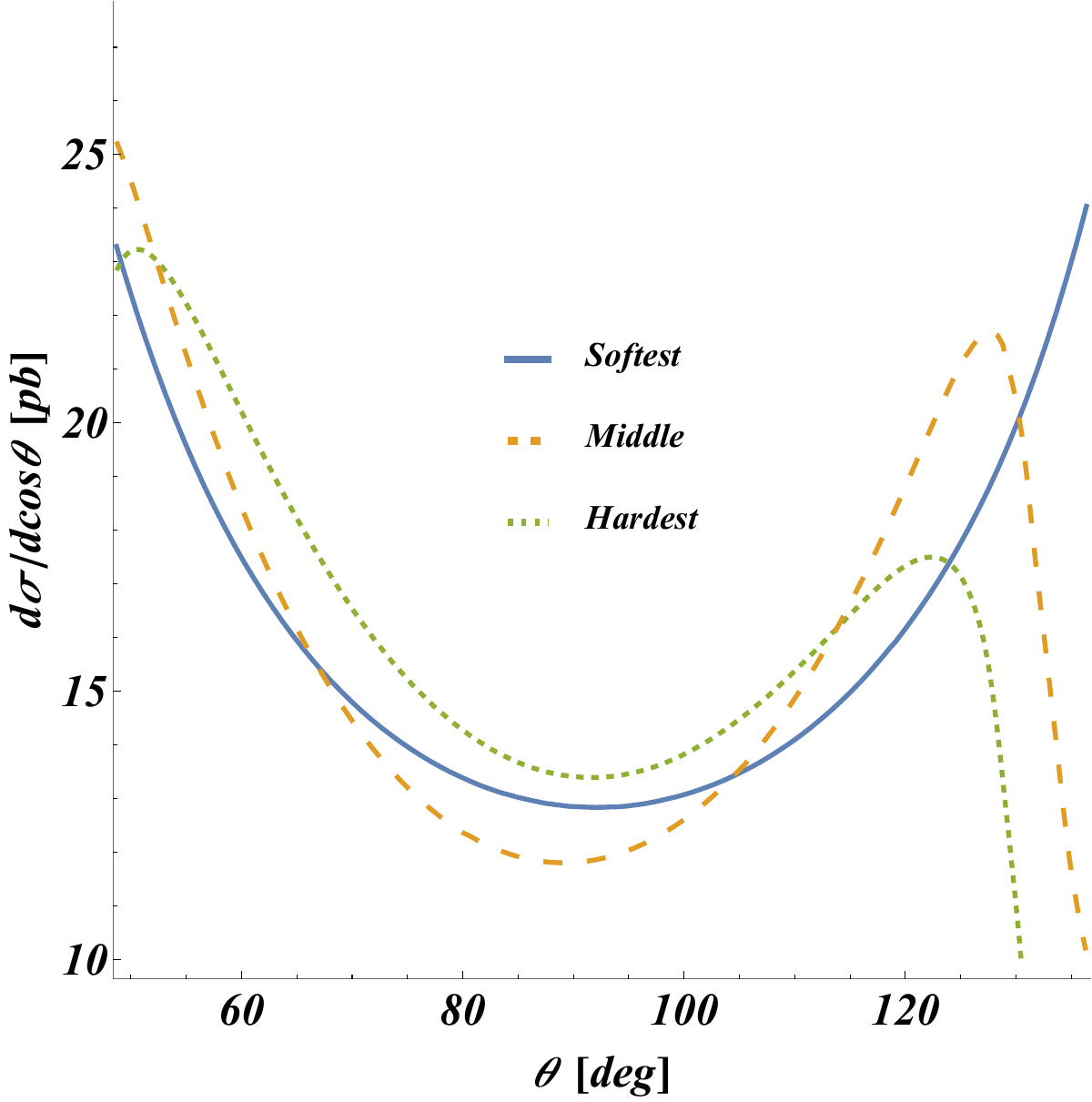} \quad \quad \includegraphics[width=0.45\textwidth]{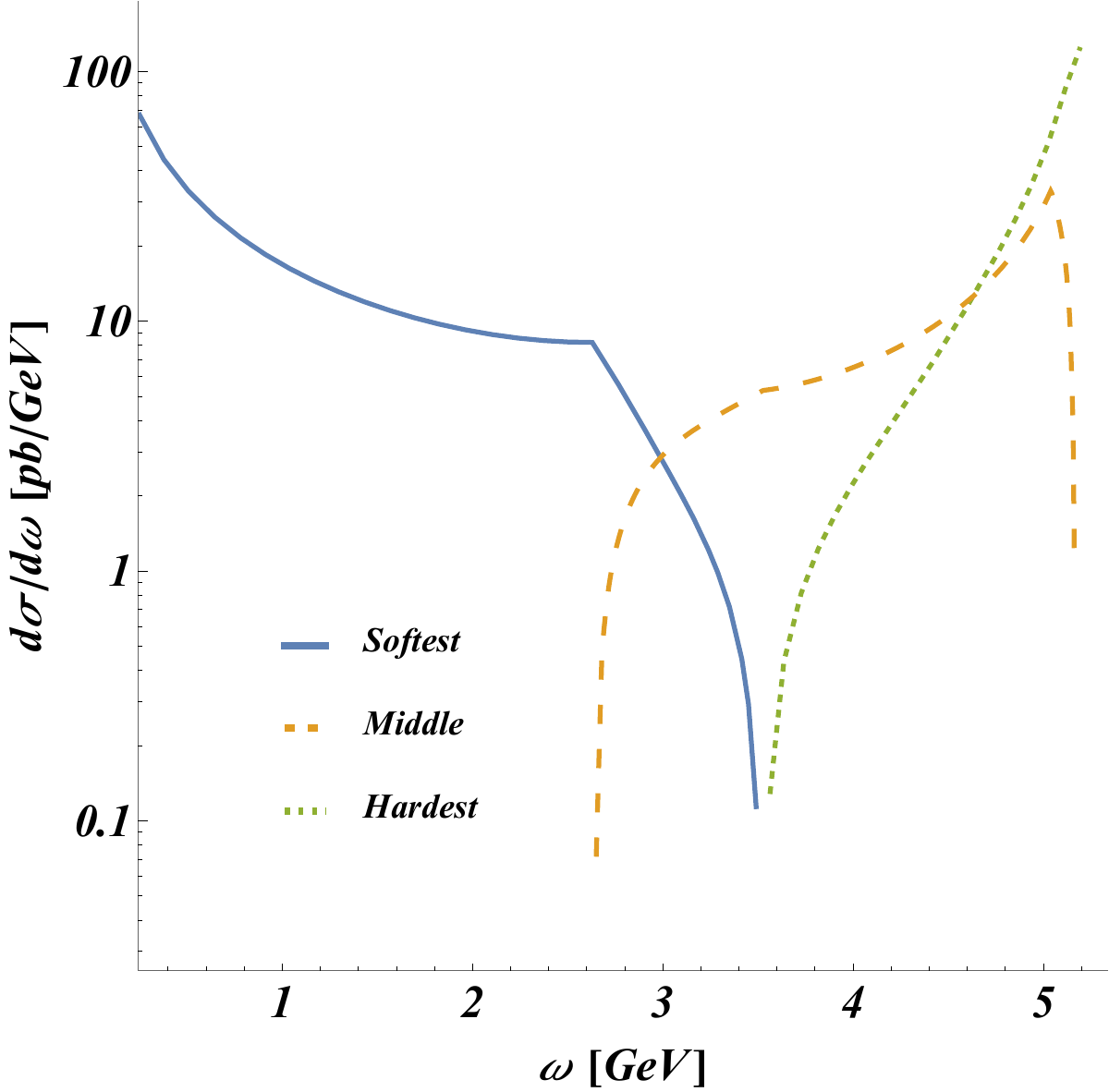} 
\caption{QED background distributions for the softest, middle and hardest photons in the $e^+e^- \to \gamma \gamma \gamma$ process in Belle II kinematics.}
\label{fig:d7}
\end{figure*}

Due to the resonant behavior of the amplitude, one photon is always emitted with a fixed energy

\begin{equation}
\omega = \frac{4E^2-m^2_a}{4E}.
\end{equation}

We note that, as the branching fraction, represented by the ratio $\Gamma_{a \gamma \gamma} / \Gamma_a$, is always less than $1$, the total cross-section of the process under investigation with an intermediate ALP may actually become smaller with a non-zero value of $g_{aee}$, compared to when this quantity is equal to zero. 

After the integration over the full phase space, the cross section of the $2 \to 3$ process with the intermediate ALP can be written in the compact form

\begin{equation}
\sigma_{e^+e^- \to \gamma\gamma \gamma} = \sigma_{e^+e^- \to a \gamma } \times \frac{\Gamma_{a \gamma \gamma}}{\Gamma_a}.
\end{equation}

If $g_{aee}=0$, the ALP decays directly to photons and this formula can be simplified further (notably, it is independent of $s$ if $m_a^2 \ll s$) as

\begin{equation}\label{eq:cs}
\sigma_{e^+e^- \to a \gamma } = \frac{\alpha g^2_{a \gamma \gamma}}{24} \left(1 - \frac{m_a^2}{s} \right)^3,
\end{equation}
where $\alpha \equiv e^2/4\pi$.

For a realistic detector, one has to integrate the cross section formula over the phase space, restricted by the experimental setup, as discussed below.

\section{Results and discussion}\label{sec3}

The ALP signal detection strategy can be based on searches for a narrow peak in the squared mass distribution $m_{\gamma \gamma}$ of photon pairs, or a narrow peak in the photon energies distributions, due to the fact that the photon which accompanies the ALP is always monoenergetic in this process.

If no significant ALP signal is observed, it is possible to constrain ALP parameters in the corresponding mass range.

In this section we illustrate our results with exclusion plots for the kinematics of an $e^+e^-$ collider. For this purpose, we first split the total $e^+e^- \to \gamma \gamma \gamma$ cross section into three terms

\begin{figure*}[t]
\centering
\includegraphics[width=0.45\textwidth]{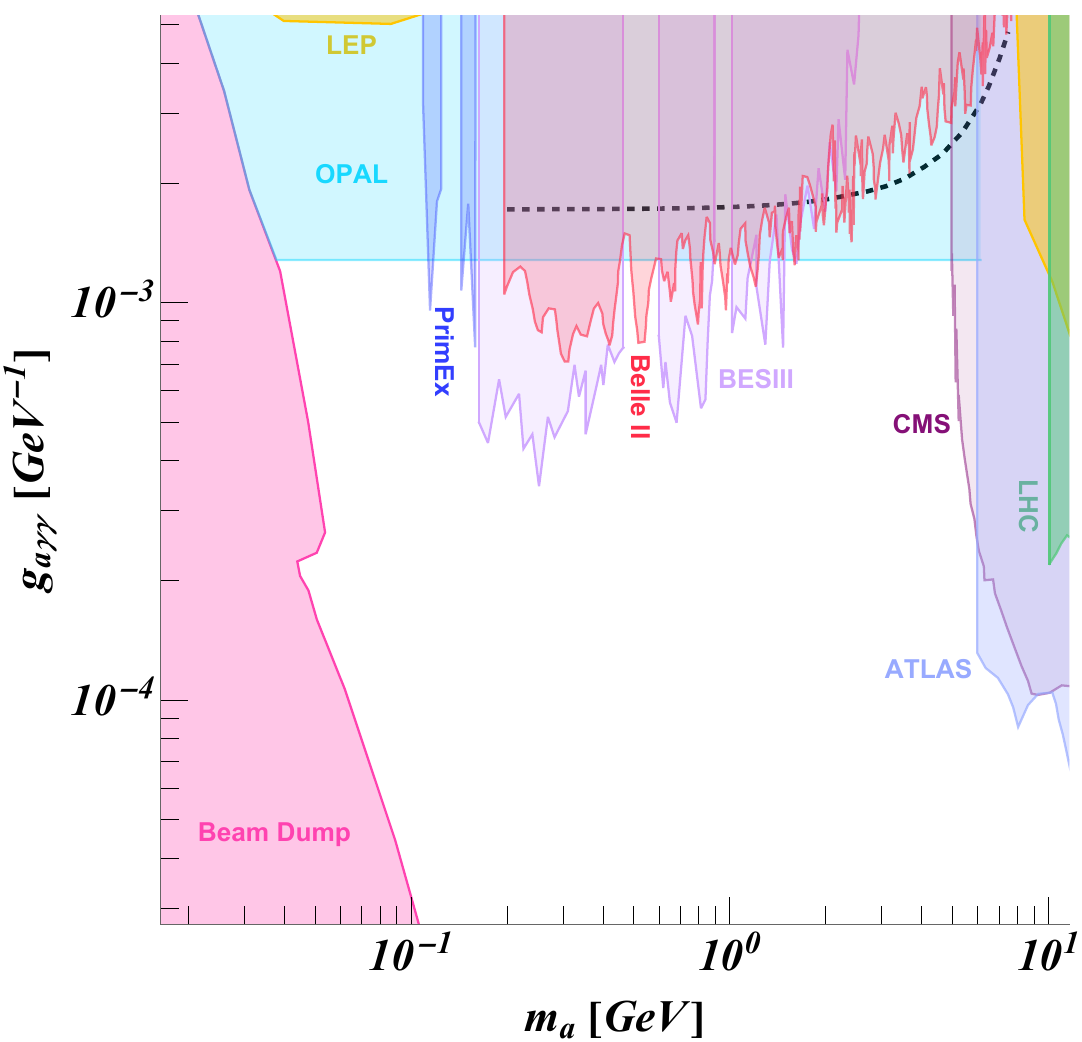} \quad \quad \includegraphics[width=0.45\textwidth]{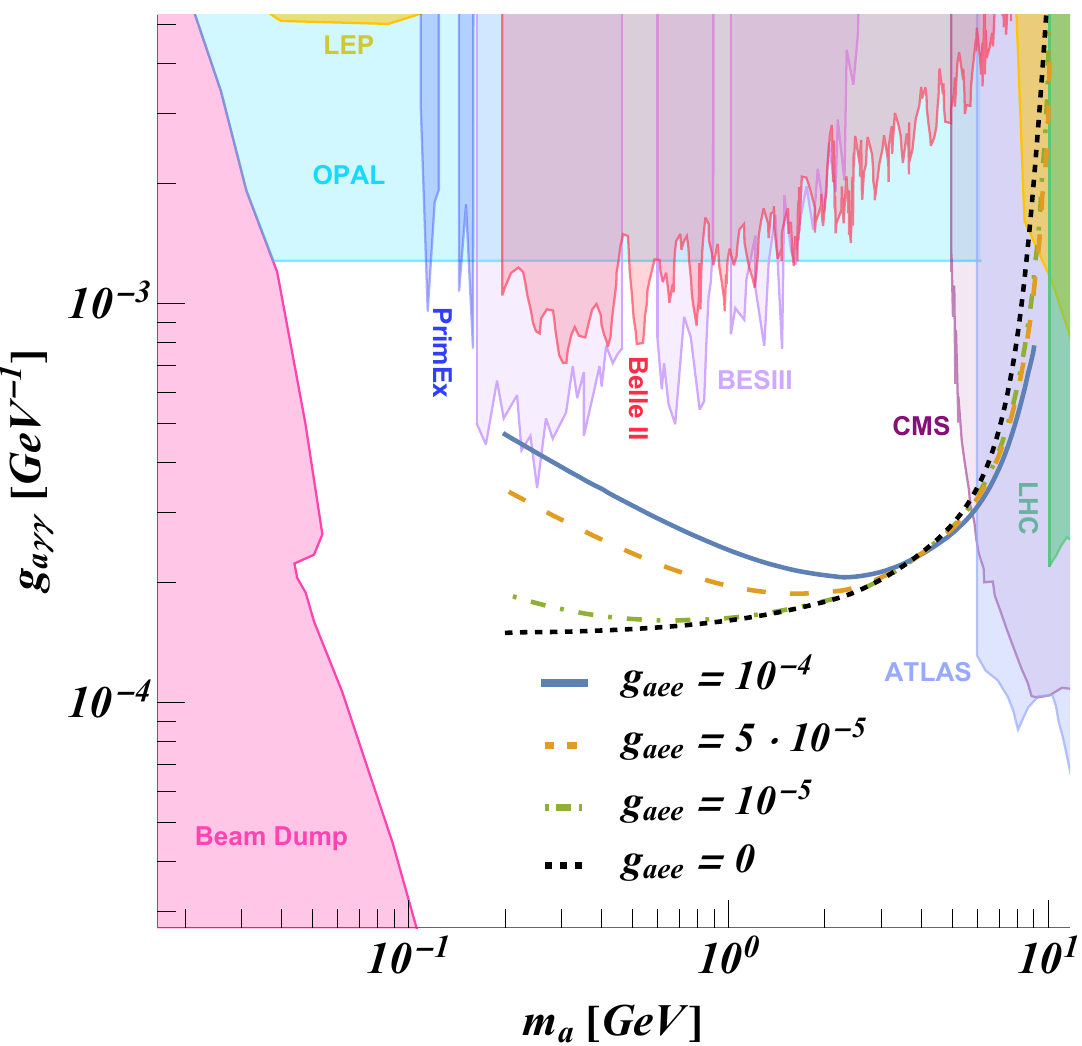} 
\caption{Left panel: Belle II constraints for $g_{a\gamma\gamma}$ based on the 2018 data set, with the analytical result shown by the black dashed curve. Right panel: projected results on the $\left(m_a, g_{a\gamma\gamma}\right)$ reach for the future data collection at Belle II corresponding to $50 \, \mbox{ab}^{-1}$ of integrated luminosity. Existing bounds (filled areas) are taken from~\cite{AxionLimits}, which in turn are based on the results of~\cite{ATLAS:2020hii,Knapen:2016moh,Aloni:2019ruo,Jaeckel:2015jla,BESIII:2022rzz,Belle-II:2020jti,CMS:2018erd,Blumlein:1990ay,NA64:2020qwq,PhysRevLett.59.755,BERGSMA1985458,Dolan:2017osp}.}
\label{fig:d8}
\end{figure*}

\begin{equation}
    \sigma_{ALP+B} = \sigma_{ALP}+\sigma_{B}+\sigma_{int},
\end{equation}
with $\sigma_{ALP}$ referring to the ALP-associated process (shown in Fig. \ref{fig:d4} and \ref{fig:d5}), while $\sigma_{B}$ is the background. The interference term $\sigma_{int}$ does not contribute since the ALP decay width $\Gamma_a$ is assumed to be much smaller than the experimental resolution of the invariant mass of the final photon pair.

The dominant part of the background originates from QED 3-photon annihilation~\cite{Dolan:2017osp}, i.e. $\sigma_{B}=\sigma_{QED}$. The aimed sensitivity is then expressed by the formula

\begin{equation}
\frac{\sigma_{ALP}}{\sigma_{QED}} = \frac{N}{\sqrt{L \cdot \sigma_{QED}}},
\end{equation}
where $L$ denotes the integrated luminosity and $N$ is the number of standard deviations that determines whether or not a fluctuation is considered as a signal. We conventionally set $N = 2$, which refers to 95\% confidence level.

In our study we neglect the potential hadronic background from $\pi^0$, $\eta$ and $\eta'$ mesons. In a complete analysis, however, those contributions must be included. Therefore, the parameter space for $m_a$ in the vicinity of the $\pi^0$, $\eta$ and $\eta'$ masses can be expected to be modified.

The $e^+e^- \to a \gamma \to \gamma \gamma \gamma$ cross section is a function of three variables. For purposes of illustration, we use the available independent constraints for $g_{aee}$ to show two-dimensional projections of $g_{a\gamma\gamma}$ as a function of $m_a$. 

Experimental searches in the MeV to GeV region are mostly focused on ALP-muon interaction~\cite{Beacham:2019nyx} and therefore not able to constrain $g_{aee}$. However, it is possible to convert constraints on visibly decaying dark photons to limits on the ALP-electron mixing~\cite{Alves:2017avw}. Indeed, the processes of $X \to e^+e^-$ and $a \to e^+e^-$ achieve comparable signal strengths in case of $g_{Xee} \sim g_{aee}$. This relation, of course, is only approximate, since the two processes have different angular distributions, but using it one can estimate $g_{aee} \lesssim  10^{-4}$~\cite{Beacham:2019nyx}.

In the following, we discuss the reach on $m_a$, $g_{aee}$ and $g_{a\gamma\gamma}$ which can be obtained from $e^+e^- \to \gamma \gamma \gamma$ data that are already available from the Belle II experiment or are expected from future running.

\subsection{Belle II kinematics}\label{subsec:31}

To obtain the exclusion plots for Belle II kinematics, we start by discussing the detector acceptance. 

Belle II is an asymmetric collider, for which electron and positron have energies of $7\, \mbox{GeV}$ and $4\, \mbox{GeV}$, respectively. This requires a boost with a relative velocity $\beta \approx 0.27$ to the center-of-momentum frame, where particles have energies of $E = 5.29 \, \mbox{GeV}$. The angular coverage of the electromagnetic calorimeter in the lab frame is $12.4^{\circ} <\theta< 155.1^{\circ}$. 

The angular region $37.3^{\circ} <\theta< 123.7^{\circ}$ provides the best energy resolution, avoiding regions close to detector gaps, and offers the lowest beam background levels~\cite{Belle-II:2020jti}. Following the work of~\cite{Dolan:2017osp}, we set the photon energy selection threshold of $0.25\, \mbox{GeV}$ in the center-of-momentum frame. Our analysis requires all three photons to be in this acceptance range and, unless otherwise specified, these experimental cuts are used for all the plots shown below.

The angular distributions for the ALP process are presented in Fig. \ref{fig:d6} for two different values of $m_a$ and two different values of $g_{aee}$. For a given $g_{aee}$, there is more than an order of magnitude difference between $m_a = 0.3 \, \mbox{GeV}$ and $m_a = 3 \, \mbox{GeV}$ curves due to the fact that for a relatively light ALP the decay width is dominated by the $a \to e^+ e^-$ channel, see Eqs. \eqref{eq:cs1} and \eqref{eq:cs2}. For the particular case of $g_{a\gamma\gamma} = 10^{-4} \, \mbox{GeV}^{-1}$ and $g_{aee} = 10^{-4}$, one obtains

\begin{align*}
&\frac{\Gamma_{a\gamma\gamma}}{\Gamma_a}  \approx 0.01, \quad \mbox{for} \quad m_a = 0.3 \, \mbox{GeV}, \\
&\frac{\Gamma_{a\gamma\gamma}}{\Gamma_a}  \approx 0.53, \quad \mbox{for} \quad m_a = 3 \, \mbox{GeV}.
\end{align*}

\subsection{QED background}\label{subsec:32}

We next discuss the QED background process. The cross section of leading order QED $e^+e^-$ annihilation in 3 photons in the massless electron limit is given by~\cite{Eidelman:1978rw}

\begin{equation}
\begin{split}
&  \bar{\sum_i} \sum_f \left|M_{e^+e^- \to \gamma\gamma\gamma \, \left(QED\right)}\right|^2 =  s \left(4\pi \alpha\right)^3 \\
& \times \frac{\sum_{i=1}^3\left(p_+k_i\right)\left(p_-k_i\right)\left[\left(p_+k_i\right)^2+\left(p_-k_i\right)^2\right]}{\prod_{i=1}^{3}\left(p_+k_i\right)\left(p_-k_i\right)}.
\end{split}
\end{equation}

For the total cross section an additional factor $1/3!$ must be added due to the 3 identical bosons in the final state. 

Fig. \ref{fig:d7} shows the corresponding QED background angular and energy distributions. In contrast to the ALP-related process (see Fig. \ref{fig:d6}), which exhibits a rather uniform angular distribution, the QED three-photon annihilation is characterized by an enhanced angular distribution in both the forward and backward directions. The presence of a distinct peak in the photon energy distribution would serve as an indication of ALP creation.

\subsection{Belle II results from 2018 data set}\label{subsec:33}

In the 2018 data run Belle II achieved an integrated luminosity of $445 \, \mbox{pb}^{-1}$~\cite{Belle-II:2020jti}, which was used for the ALP searches in a simplified way by converting the cross section limit to the coupling using Eq. \eqref{eq:cs}. The latter formula does not take into consideration the fact that all three photons in the ALP-associated process must be detected in the acceptance range of the electromagnetic calorimeter.

We require three resolved photons with energies higher than $0.65 \, \mbox{GeV}$ in the center-of-momentum frame as a criteria for this event selection. These requirements are slightly different from those used in the Belle II report~\cite{Belle-II:2020jti}, where the selection of photons with energies above $0.65 \, \mbox{GeV}$ (for $m_a > 4 \, \mbox{GeV}$) and $1 \, \mbox{GeV}$ (for $m_a \leq 4  \,\mbox{GeV}$) in the lab frame was performed. The difference is negligible since $g_{a \gamma \gamma}$ is sensitive to $\sigma_{QED}^{-1/4}$.

\begin{figure}[t]
\centering
\includegraphics[width=0.45\textwidth]{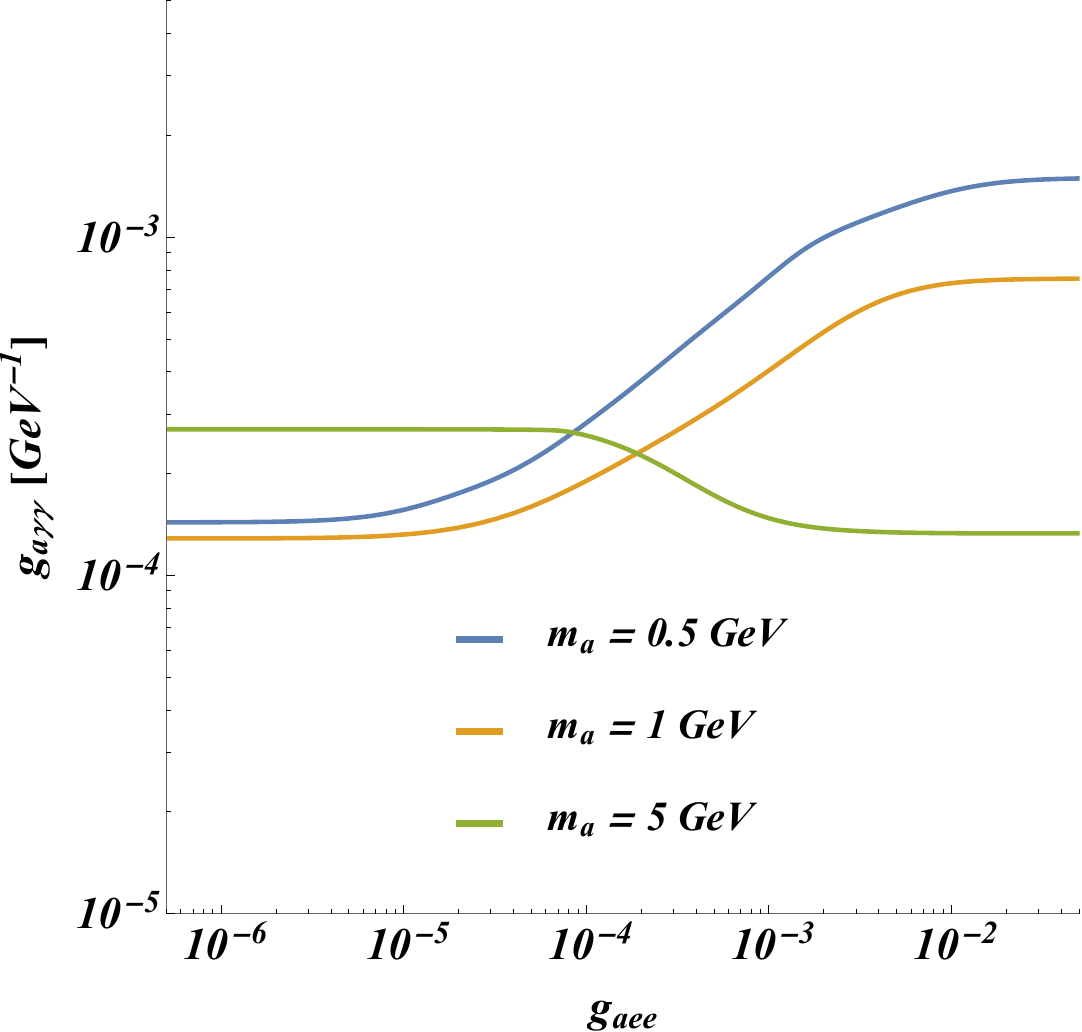}
\caption{Projected results on the $\left(g_{aee}, g_{a\gamma\gamma}\right)$ reach for three different $m_a$ hypotheses for the future data collection at Belle II corresponding to $50 \, \mbox{ab}^{-1}$ of integrated luminosity. Existing bounds, based on the Belle II 2018 run, can be found in~\cite{Liu:2023bby}.}
\label{fig:d9}
\end{figure}

Our result based on Eq. \eqref{eq:cs} is shown on Fig. \ref{fig:d8} (left panel) by the black curve. It shows a good agreement with the analysis of~\cite{Belle-II:2020jti} in the higher ALP mass region. In the lower mass region some deviations are seen. This is expected because in the case of a light ALP the invariant mass of a photon pair also becomes low, i.e. two photons travel in a very narrow cone with each other, oppositely to the third photon. This produces very asymmetric kinematics and the QED background becomes suppressed. In our analysis we do not take this into consideration, but more detailed investigation can be performed in future work. 

\subsection{Belle II projection from upcoming data collection}\label{subsec:34}

Belle II is expected to reach an integrated luminosity of $50 \, \mbox{ab}^{-1}$ after around 10 years of running. The resulting constraints which such future data would yield were investigated in~\cite{Dolan:2017osp} for the case where ALPs are only coupled to photons (i.e. for $g_{aee} = 0$). Fig. \ref{fig:d8} right panel shows the projected sensitivity in two scenarios: ALPs coupled only to photons and ALPs coupled to photons and electrons with different $g_{aee}$ coupling strength.

Our results for the scenario $g_{aee} = 0$ are in reasonably good agreement with the exclusion limits deduced in~\cite{Dolan:2017osp} in the high $m_a$ region. For lower values of $m_a$ one expects a similar deviation as for the 2018 Belle II data discussed above. Fig. \ref{fig:d8} also shows that the inclusion of a non-zero interaction of ALPs with electrons significantly affects the final result, especially in ALP mass range $m_a \lesssim 2 \, \mbox{GeV}$. The assumption $g_{aee}=0$ generally leads to an overestimated $g_{a\gamma\gamma}$ limit, which may be incorrect if the ALP has other decay channels besides the 2-photon mode.

Notably, such an interplay of two couplings becomes relevant only with higher statistics and does not seriously affect Belle II 2018 results. Indeed, at $m_a \sim 1 \, \mbox{GeV}$,  $g_{aee} \sim 10^{-4}$ and  $g_{a\gamma\gamma} \gtrsim 10^{-3} \, \mbox{GeV}^{-1}$ the branching ratio can be estimated as:

\begin{equation*}
\frac{\Gamma_{a\gamma\gamma}}{\Gamma_a} \gtrsim 0.9,
\end{equation*}
i.e. the decay into electron-positron pair is largely suppressed. However, with the several orders of magnitude increase in integrated luminosity, the reach on the $\left(m_a,g_{a\gamma\gamma}\right)$ is considerably improved and the opposite becomes true (see the end of the Section \ref{subsec:31} for more details). In such case, more detailed models with additional parameters are required to constrain invisible particles in a more rigorous way. 

Fig. \ref{fig:d9} shows the reach on $\left(g_{aee}, g_{a\gamma\gamma}\right)$ for three different values of $m_a$, which can be achieved with the Belle II upcoming data collection.

\section{Conclusion}\label{sec4}

In this paper we discussed ALPs coupled to electrons and photons in a minimal way. The contributions of ALP states to 2- and 3-photon $e^+e^-$ annihilation events were calculated. In this way, we obtained new constraints for possible ALPs in the MeV to GeV mass range, which can be tested at $e^+e^-$ colliders. Results were shown for Belle II kinematics both from existing data and from forthcoming data with projected integrated luminosity of $50 \, \text{ab}^{-1}$.

Our results indicate that the $g_{a \gamma \gamma}$ limits can be vastly affected in the presence of an additional decay mode, especially in the lower $m_a$ region. The difference between single $g_{a \gamma \gamma}$ and combined $g_{a \gamma \gamma}$/$g_{aee}$ scenarios is minimal for Belle II 2018 data, but is expected to become substantial with the data currently being collected. Consequently, the constraints on lower mass ALP will be less stringent.

Using current best limits for $g_{aee}$, it is possible to improve the $g_{a \gamma \gamma}$ limits by at least an order of magnitude, which allows to significantly narrow down the search area for potential ALPs in the MeV to GeV mass range. This result can be improved further if a better way to constrain $g_{aee}$ independently is available. 

There are many possible ways to further extend this work. First of all, a more precise background modeling and experimental analysis must be performed in order to refine the exclusion plots, especially around the $\pi^0$, $\eta$ and $\eta'$ masses. The comparison with the Belle II 2018 data analysis shows that with more detailed background analysis it is possible to even further improve $g_{a \gamma \gamma}$ constraints in the $m_a^2 \ll s $ case.

Additionally, in this paper we assumed that ALPs interact only with the electrons and photons. Hidden decay channels were not considered. However, the contribution of light dark matter particles of sub-GeV masses may change the obtained constraints if the pair production threshold is surpassed. At the same time, the inclusion of such particles makes it more complicated to set any constraints, as new free parameters appear.

Finally, we restricted ourselves with ALPs which are not coupled to muons (and taus). However, lepton universality leads to an increase in the coupling constant $g_{a\mu\mu}$ by around two orders of magnitude compared to $g_{aee}$. It may notably affect the results if $m_a$ is larger than $2m_{\mu}$. The parameter space in such case will get additional restrictions from the requirement of compatibility with current $\left(g-2\right)_{\mu}$ data~\cite{Buen-Abad:2021fwq,Muong-2:2023cdq}.

\bmhead{Acknowledgments}

This work was supported by the Deutsche Forschungsgemeinschaft (DFG, German Research Foundation), in part through the Research Unit [Photon-photon interactions in the Standard Model and beyond, Projektnummer 458854507 - FOR 5327], and in part through the Cluster of Excellence [Precision Physics, Fundamental Interactions, and Structure of Matter] (PRISMA$^+$ EXC 2118/1) within the German Excellence Strategy (Project ID 39083149).

\appendix
\section{Loop-induced effects}\label{appendix:a}

The correction $\delta g_{a\gamma\gamma}$ to the tree-level ALP-photon coupling  $g^0_{a\gamma\gamma}$  induced by the electron loop must be considered in two distinct scenarios (in contrast to the ALP-electron coupling which is unambiguous as it is represented only once in each of the relevant diagrams). In the case of the Primakoff-like process with one photon being real and one - virtual, carrying the momentum $q$, the corresponding contribution is given by~\cite{Ferreira:2022xlw} 

\begin{equation}
\delta g^{\text{Primakoff}}_{a\gamma\gamma}  = \frac{\alpha g_{aee}^0}{\pi m_e} \left[1 + F\left(m_a^2,q^2\right)\right],
\end{equation}
the function $F\left(m_a^2,q^2\right)$ is of the form

\begin{equation}
\begin{split}
& F\left(m_a^2,q^2\right) = \left(\frac{4m_e^2}{m_a^2}\right)\left( \frac{m_a^2}{m_a^2 - q^2}\right)  \\
& \cdot \left[f^2\left(\frac{4m_e^2}{q^2}\right)-f^2\left(\frac{4m_e^2}{m_a^2}\right)\right], \\
\end{split} 
\end{equation}
whereas $f\left(x\right)$ is defined by

\begin{equation}
\begin{split}
& f\left(x\right) = \Theta\left(x-1\right) \arcsin{\left(\frac{1}{\sqrt{x}}\right)} \\
& + \frac{1}{2} \Theta\left(1-x\right) \left[\pi+i \ln{\left(\frac{1+\sqrt{1-x}}{1-\sqrt{1-x}}\right)}\right],
\end{split}
\end{equation}
with $\Theta\left(x\right)$ being the step function.

Oppositely, for the ALP decay into two real photons one has $q^2=0$, leading to:

\begin{align}
& \delta g^{\text{decay}}_{a\gamma\gamma} = \frac{\alpha g_{aee}^0}{\pi m_e} \left[1 + F\left(m_a^2,0\right)\right], \\
& F\left(m_a^2,0\right) = -\left(\frac{4m_e^2}{m_a^2}\right) \cdot f^2\left(\frac{4m_e^2}{m_a^2}\right).
\end{align}

\begin{figure}[t]
\centering
\includegraphics[width=0.45\textwidth]{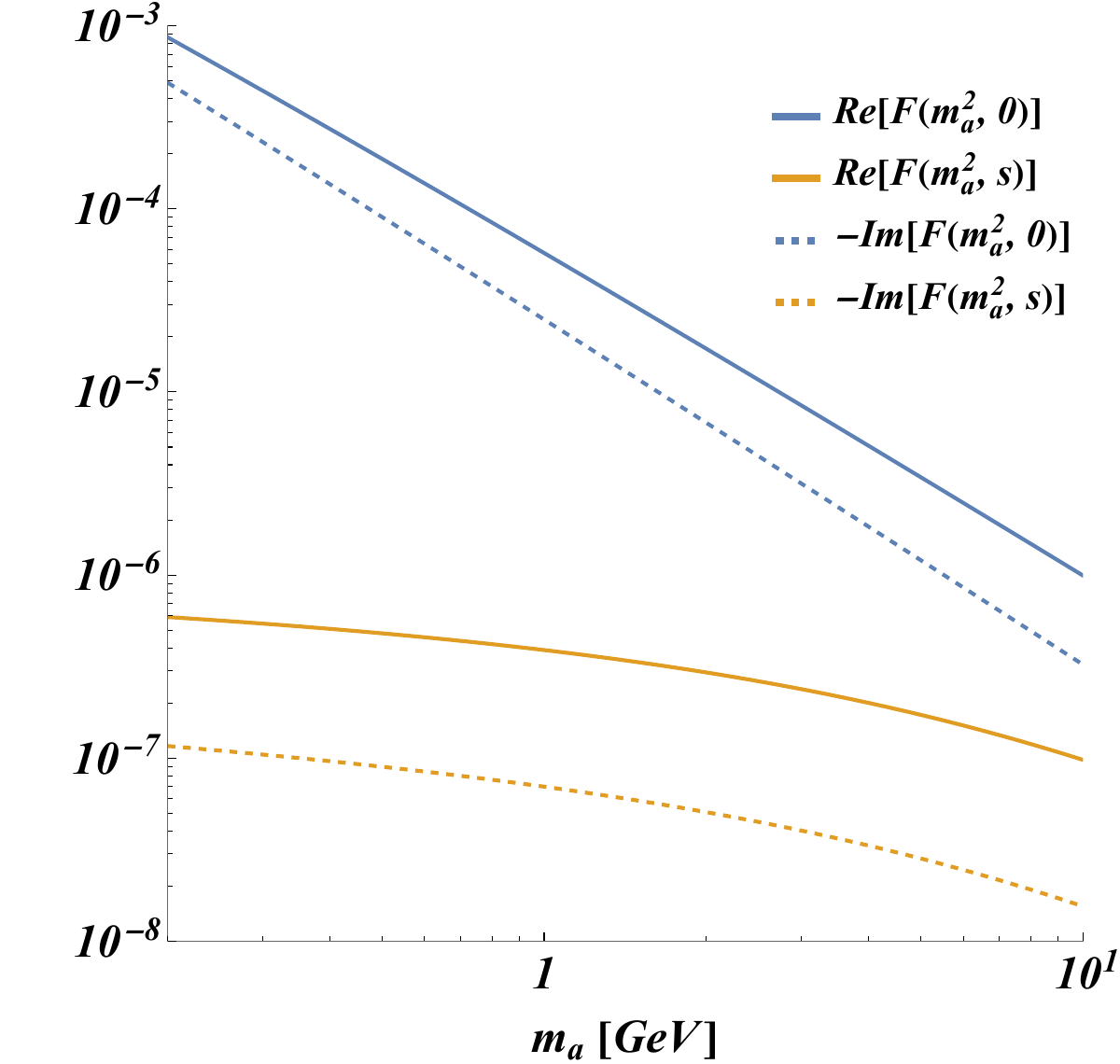}
\caption{Real and imaginary components of $F\left(m_a^2,0\right)$ and $F\left(m_a^2,q^2\right)$, which distinguish $g^{\text{decay}}_{a\gamma\gamma}$ and $g^{\text{Primakoff}}_{a\gamma\gamma}$, as functions of $m_a$. The value $s= 112 \, \mbox{GeV}^2$ is set, corresponding to the kinematics of the Belle II experiment.}
\label{fig:d10}
\end{figure}

Both $F\left(m_a^2,q^2\right)$ and $F\left(m_a^2,0\right)$ are suppressed by the factor of $4m_e^2/m_a^2$. However, due to its structure, $F\left(m_a^2,q^2\right)$ appears to be suppressed twice

\begin{equation}
\begin{split}
& m_a^2\frac{f^2\left(\frac{4m_e^2}{q^2}\right)-f^2\left(\frac{4m_e^2}{m_a^2}\right)}{m_a^2-q^2} \\
&\left. \approx  \left(\frac{4m_e^2}{q^2}\right) \frac{df^2 (x) }{d x}\right|_{x = \frac{4m_e^2}{m_a^2}} + O\left(\frac{4 m_e^2}{m_a^2}\left(\frac{q^2-m_a^2}{q^2}\right)\right).
\end{split}
\end{equation}

Fig. \ref{fig:d10} illustrates the behaviour of $F\left(m_a^2,q^2\right)$ and $F\left(m_a^2,0\right)$ within the relevant domain, specifically Belle II kinematics where $\sqrt{s} = 10.58 \, \mbox{GeV}$. We see that both functions approach zero as $m_a$ increases and bring the correction roughly of

\begin{gather}
\delta g^{\text{decay}}_{a\gamma\gamma} = \frac{\alpha g_{aee}^0}{\pi m_e} \left[1 + O\left(10^{-3}\right)\right], \\
\delta g^{\text{Prima}}_{a\gamma\gamma} = \frac{\alpha g_{aee}^0}{\pi m_e} \left[1 + O\left(10^{-6}\right)\right],
\end{gather}
clearly illustrating that for the Belle II analysis the difference of two effective ALP-photon couplings is negligible. 

The correction itself is enhanced by the factor $1/m_e$ and can be estimated as

\begin{equation}
\delta g_{a\gamma\gamma} \approx \frac{4.6 g_{aee}^0}{\text{GeV}} \lesssim  \frac{4.6 \cdot 10^{-4}}{\text{GeV}},
\end{equation}
which, however, does not exclude any parameter space by itself since the experiment is only sensitive to the effective coupling $ g^{\text{eff}}_{a\gamma\gamma} =g^{0}_{a\gamma\gamma} + \delta g_{a\gamma\gamma}$.

\bibliographystyle{sn-mathphys} 
\bibliography{sn-bibliography}% common bib file

\end{document}